\documentclass[journal]{IEEEtran}
\usepackage{algpseudocode}

\usepackage{graphicx}
\usepackage[]{algpseudocode}
\usepackage{algorithmicx,algorithm}
\usepackage{amsmath}
\usepackage{titlesec}
\usepackage{multirow}  
\usepackage{amsthm,amsmath,amssymb}
\usepackage{graphicx}
\usepackage{float}
\usepackage{subfigure}
\usepackage{mathrsfs}
\usepackage{url}
\usepackage{array}
\newcolumntype{C}[1]{>{\centering}p{#1}}
\usepackage{makecell}
\usepackage{diagbox}
\usepackage{amsmath}
\usepackage{gensymb}
\usepackage{cite}
\usepackage{graphicx}
\usepackage{subfigure}
\usepackage{color}

\ifCLASSOPTIONcompsoc
 \usepackage[caption=false,font=normalsize,labelfont=sf,textfont=sf]{subfig}
\else
 \usepackage[caption=false,font=footnotesize]{subfig}
\fi

\UseRawInputEncoding
\begin{document}
\vspace{-15mm}
\title{RIS Design for CRB Optimization in Source Localization with Electromagnetic Interference}
 
\author{Yuhua Jiang, Yuanwan Mai, and Feifei Gao 

\thanks{Y. Jiang, and F. Gao are with Institute for Artificial Intelligence, Tsinghua University (THUAI), 
State Key Lab of Intelligent Technologies and Systems, Tsinghua University, 
Beijing National Research Center for Information Science and Technology (BNRist), Beijing, P.R. China (email:  jiangyh20@mails.tsinghua.edu.cn, feifeigao@ieee.org).



W. Yuan is with Institute Information Science Academy of CETC (email: yuanwanmai7@163.com).

}
}

\maketitle
\vspace{-15mm}
\begin{abstract}
Reconfigurable Intelligent Surface (RIS) plays an
important role in enhancing source localization accuracy.
Based on the information inequality of Fisher information analyses, the Cram\'{e}r-Rao Bound (CRB) of the localization error can be used to evaluate the localization accuracy for a given set of RIS coefficients. 
In this paper, we adopt the manifold optimization method to derive the optimal RIS coefficients that minimize the CRB of the localization error with the presence of electromagnetic interference (EMI), where the RIS coefficients are restricted to lie on the complex circle manifold. 
Simulation results are provided to validate the proposed studies under various circumstances. 

\end{abstract}

\IEEEpeerreviewmaketitle

\section{Introduction} 
Recently, the reconfigurable intelligent surface (RIS) has been used in integrated sensing and communications systems \cite{mypaper}.
RIS plays an essential role in source localization when the line-of-sight (LOS) path between the transmitter and the receiver is blocked, hence making localization feasible when the conventional technologies fail. 
Moreover, RIS is beneficial to timely improve the localization accuracy when
the LoS path is present. Such flexibility makes RIS a pivotal localizing technology.

As for RISs-based localization, directional
reflection beam design has been considered with 
prior knowledge of the user equipment (UE) location, which aims to concentrate the reflected power towards the UE and thus increases the localization accuracy \cite{9}.
Alternatively, the authors in \cite{10} use simpler random RIS phase profiles for
asynchronous positioning in downlink single-input-single-output (SISO) transmission. This scheme does not require any prior information, but cannot provide the optimal RIS design
under a priori UE location information. In \cite{11}, the localization in narrow-band systems is extended to wideband systems with an unfavorable assumption that the phase responses of RIS elements must be constants over the considered frequency band.
In wideband localization, RIS can be designed by dividing the frequency
band into a sum of narrow frequency bands \cite{12}.

Based on the information inequality in Fisher information analyses, the Cram\'{e}r-Rao Bound (CRB) can be used
to evaluate the localization accuracy for a given set of RIS coefficients. 
 In \cite{NLN}, the electromagnetic (EM) models are built to compute CRB of localization methods for both discrete and continuous RISs.
In \cite{1}, the RIS coefficients are designed to focus energy on one of the anchors, which yields lower CRB than that with randomly designed RIS coefficients.
In \cite{PEB},  CRB is derived for both position and orientation error of localizing a rotated user equipment. 

However, the CRB minimization problems in \cite{NLN}-\cite{PEB} are solved without considering the electromagnetic interference (EMI), which is inevitably caused by various uncontrolled electromagnetic
sources such as natural radiation, man made devices and is 
also reflected by the RIS towards the user \cite{EMI}. To combat EMI, RISs need to be designed in a modified way compared with EMI-free scenarios.
To our best knowledge, no RIS optimization method has been proposed to minimize CRB under the influence of EMI.



In this paper, we apply the manifold optimization method to derive the optimal CRB of the localization error with EMI under practical RIS hardware limitations (e.g., unit-modulus values with quantized phases). In order to find the iterative search direction, we provide the closed-form Wirtinger derivatives in the gradient descent part of the optimization method. 
Simulation results show that the proposed method can significantly decrease the CRB of the localization error with the presence of EMI.


\section{System Model}
Consider a three-dimensional (3D) scenario with one RIS, one agent that acts as the source, and $M$ anchors that act as receivers, as shown in Fig.~1. 
Denote the coordinate of the 
agent and the $m$th anchor by $\boldsymbol{q}=[q_1,q_2,q_3]^T$ and $\boldsymbol{p}_m=[x_m,y_m,z_m]^T$, respectively. The position of the agent is unknown, while the positions of the anchors are known.
We consider the RIS as a rectangular plate with length $a$ in $y$-axis and length $b$ in $x$-axis, located in the horizontal plane. 
Suppose the RIS is equipped with $N$ passive elements, and each element is $l_1$ long and  $l_2$ wide.
Suppose the wave number is $k_0=2\pi/\lambda_0$, where $\lambda_0$ is the wavelength of the transmitted electromagnetic waves.

Denote $r_m$ as the distance from the $m$th anchor to the agent, $\rho_n$ as the distance from the $n$th element of the RIS to the agent, and $d_{mn}$ as the distance from the $m$th anchor to the $n$th element of the RIS.
The overall channel vector from the transmitter to RIS and RIS to the $m$th anchor is denoted as $\boldsymbol{h}_{RIS(m)}$ $(m=1,\cdots,M)$. 
The direct path channel from the transmitter to the $m$th anchor is denoted as $h_{DP(m)}$ $(m=1,\cdots,M)$. 
Define the  steering vectors $\boldsymbol{a}_{m,1}=\left[e^{-jk_0\rho_1},\cdots,e^{-jk_0\rho_N} \right]^T$ and $\boldsymbol{a}_{m,2}=\left[e^{-jk_0d_{m1}},\cdots,e^{-jk_0  d_{mN} } \right]^T$.
The two channels can be respectively written as \cite{1}, \cite{NLN}
\begin{align}
\boldsymbol{h}_{RIS(m)}&= \alpha_m
\boldsymbol{a}_{m,1} \odot \boldsymbol{a}_{m,2} , \\
h_{DP(m)}&=\beta_m e^{-j k_0 r_m},
\end{align}
where $\alpha_m$ and $\beta_m$ are pathloss for the $m$th scattering path and the direct path, respectively. Define $\boldsymbol{\gamma}=[\alpha_1,\cdots,\alpha_M,\beta_1,\cdots,\beta_M]^T$ and $\tilde{\boldsymbol{\gamma}}=[\Re\{\boldsymbol{\gamma}\}^T,\Im\{\boldsymbol{\gamma}\}^T]^T$.
Denote $[t_n,u_n,0]^T$ as the location of the $n$th element of the RIS, and there are
\begin{align}
\rho_n&=\sqrt{(q_1-t_n)^2+(q_2-u_n)^2+q_3^2}, \\
r_m&=\sqrt{(q_1-x_m)^2+(q_2-y_m)^2+(q_3-z_m)^2}.
\end{align}

Let $\boldsymbol{w}=[w_1,\cdots,w_N]^T$ be the vector containing the reflection coefficients of the RIS to be designed.
Since the passive elements on the RIS can not adjust the amplitude of the incident EM waves, there is $|w_n|=1, \forall n$. 
Denote  $\boldsymbol{h}_{m,2}=\zeta_m \boldsymbol{a}_{m,2}$ as the channel  from RIS to the $m$th anchor and $\boldsymbol{n}_{EMI}$ as the incident EMI field on the surface of the RIS. Denote $\boldsymbol{H}_{m,2}=\operatorname{diag\{\boldsymbol{h}_{m,2}\}}$.
Then, the $t$th signal received by the $m$th anchor is 
\begin{equation}
[\boldsymbol{y}_m]_t=\underbrace{(\boldsymbol{h}_{RIS(m)}^T\boldsymbol{w}+h_{DP(m)})[\boldsymbol{x}]_t}_{[\boldsymbol{s}_m]_t}+
(\boldsymbol{H}_{m,2} \boldsymbol{w})^T\boldsymbol{n}_{EMI}+n_m,
\end{equation}
where $[\boldsymbol{y}_m]_t$ denotes the $t$th element of the vector $\boldsymbol{y}_m$,
$\boldsymbol{x}$ denotes the $T$ probing signals transmitted by the agent, $\boldsymbol{s}_{m}$ denotes the received signal of the $m$th anchor without noise, and $\boldsymbol{n}_m \sim \mathcal{N}_{\mathbb{C}^T}\left(0, \sigma^2_m\right)$ denotes the thermal noise disturbing the signal reception at the $m$th anchor. 


\subsection{Electromagnetic Interference Modeling}
Suppose the EMI is produced by external sources located in the halfspace in front of the RIS. Let $E_{EMI}$ denote the magnitude of the energy flux density of the EMI at the RIS.
The statistical interference correlation matrix $\mathbb{E}\left\{\boldsymbol{n}_{EMI} \boldsymbol{n}_{EMI}^{\mathrm{H}}\right\} \triangleq \boldsymbol{R}$ is assumed known. When EMI is uniformly distributed from all angles, $\boldsymbol{R}$ can be formulated as 
\begin{equation}
\left[\mathbf{R}\right]_{n, m}=E_{EMI}\operatorname{sinc}\left(\frac{2\sqrt{(t_n-t_m)^2+(u_n-u_m)^2}}{\lambda}\right),
\end{equation}
where $\operatorname{sinc}(x)=\frac{\sin(\pi x)}{\pi x}$.
Note that EMI is reasonably treated  as noise because it is generated by uncontrollable signals. The general noise power at the $m$th anchor is 
\begin{align}
P_m&=\mathbb{E}\left\{\left( |(\boldsymbol{H}_{m,2} \boldsymbol{w})^T\boldsymbol{n}_{EMI}|^2+|n_m|^2\right)\right\}\nonumber\\
&= \boldsymbol{w}^T \boldsymbol{H}_{m,2}^T \boldsymbol{R} \boldsymbol{H}_{m,2}^* \boldsymbol{w}^*+\sigma^2_m.
\end{align}

After receiving signals from the agent, the anchors can jointly estimate the location of the 
agent by methods such as the maximum likelihood estimator (MLE) \cite{PEB}.
\begin{figure}[t]
  \centering
  \centerline{\includegraphics[width=6.7cm,height=5cm]{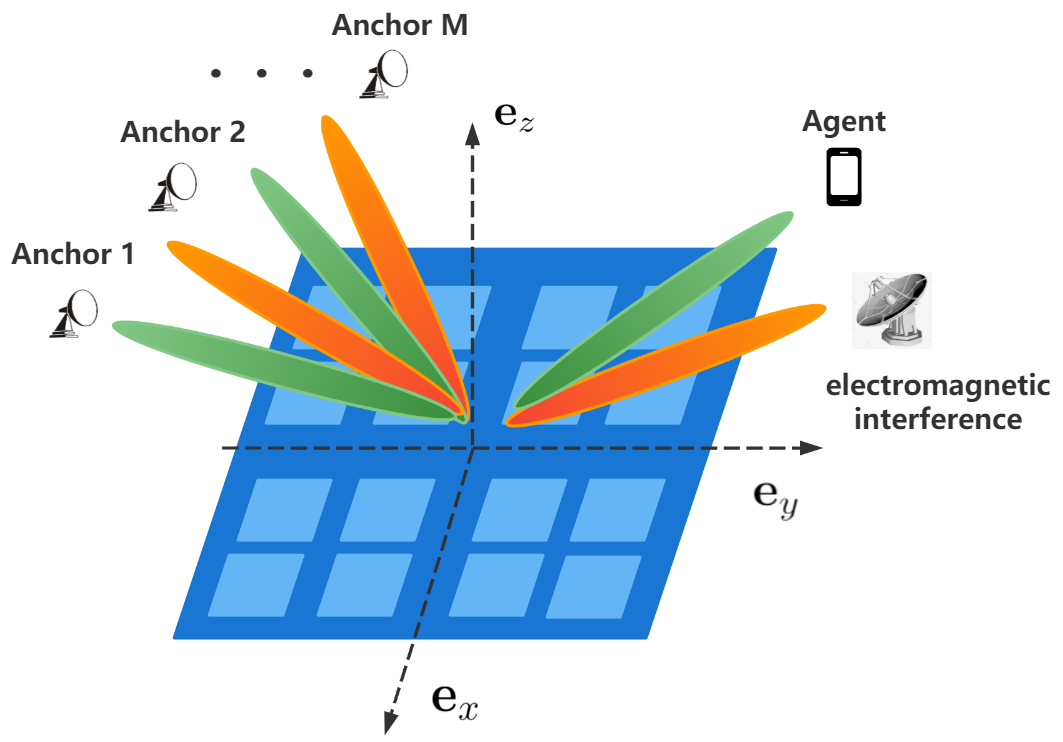}}
  \caption{The localization system with one agent, $M$ anchors, and one RIS.}
  \label{image}
\end{figure}

\section{The Cram\'{e}r-Rao Bound of Localization Error}
Denote the Fisher information matrix (FIM) of the $m$th anchor as $\boldsymbol{J}_m(\boldsymbol{q})$, $(m=1,\cdots,M)$. 
According to \cite{3}, the general FIM $\boldsymbol{J}(\boldsymbol{q})$ can be expressed  as 
\begin{align}
\boldsymbol{J}(\boldsymbol{q})=\sum_{m=1}^M \boldsymbol{J}_m(\boldsymbol{q}).
\end{align}
Let $\boldsymbol{\eta}=[\boldsymbol{q}^T,\tilde{\boldsymbol{\gamma}}^T]^T$ collect all unknown parameters. 
The elements of $\boldsymbol{J}_m(\boldsymbol{q})$ are given by
\begin{align}
[\boldsymbol{J}_m(\boldsymbol{q})]_{i, k}=\frac{2}{P_m} \Re\left\{
\frac{\partial \boldsymbol{s}_{m}^{H}}{\partial \eta_{i}} 
\frac{\partial \boldsymbol{s}_{m}}{\partial \eta_{k}}\right\}.
\label{FIM}
\end{align} 
As a consequence, $\boldsymbol{J}_m(\boldsymbol{q})$ can be partitioned as
\begin{equation}
\boldsymbol{J}_m(\boldsymbol{q})=\left[\begin{array}{ll}
\boldsymbol{J}_{\boldsymbol{q}\boldsymbol{q},m} & \boldsymbol{J}_{\boldsymbol{q}\tilde{\boldsymbol{\gamma}},m} \\
\boldsymbol{J}_{\boldsymbol{q}\tilde{\boldsymbol{\gamma}},m}^T & \boldsymbol{J}_{\tilde{\boldsymbol{\gamma}} \tilde{\boldsymbol{\gamma}},m}
\end{array}\right].
\end{equation}
The overall FIM $\boldsymbol{J}(\boldsymbol{q})$ can then be written as 
\begin{equation}
\boldsymbol{J}(\boldsymbol{q})=\sum_{m=1}^{M}\left[\begin{array}{ll}
\boldsymbol{J}_{\boldsymbol{q}\boldsymbol{q},m} & \boldsymbol{J}_{\boldsymbol{q}\tilde{\boldsymbol{\gamma}},m} \\
\boldsymbol{J}_{\boldsymbol{q}\tilde{\boldsymbol{\gamma}},m}^T & \boldsymbol{J}_{\tilde{\boldsymbol{\gamma}} \tilde{\boldsymbol{\gamma}},m}
\end{array}\right]
\overset{\Delta}{=}
\left[\begin{array}{ll}
\boldsymbol{J}_{\boldsymbol{q}\boldsymbol{q}} & \boldsymbol{J}_{\boldsymbol{q}\tilde{\boldsymbol{\gamma}}} \\
\boldsymbol{J}_{\boldsymbol{q}\tilde{\boldsymbol{\gamma}}}^T & \boldsymbol{J}_{\tilde{\boldsymbol{\gamma}} \tilde{\boldsymbol{\gamma}}}
\end{array}\right].
\end{equation}

By substituting (1) and (2) into (3)  and using the chain rule of derivative, we obtain
\begin{align}
& \frac{\partial \boldsymbol{s}_{m}}{\partial q_{i}} = -j k_0 (\alpha_m
 \sum_{n=1}^N  w_n e^{-jk_0(\rho_n+d_{mn})} \frac{\partial \rho_n}{\partial q_{i}}
\nonumber\\&+ \beta_m e^{-jk_0 r_m} \frac{\partial r_m}{\partial q_{i}} ) \boldsymbol{x} 
 \triangleq  \sum_{n=1}^N w_n A_{mni}+c_{mi}, \quad i=1,2,3 \\
\label{FIM2}
& [\frac{\partial \boldsymbol{s}_{m}}{\partial \Re\{\alpha_{m'}\}},\frac{\partial \boldsymbol{s}_{m}}{\partial \Im\{\alpha_{m'}\}}] = \delta_{mm'}
 \sum_{n=1}^N  w_n e^{-jk_0(\rho_n+d_{mn})} \boldsymbol{x}\otimes  [1,j],\\
& [\frac{\partial \boldsymbol{s}_{m}}{\partial \Re\{\beta_{m'}\}},\frac{\partial \boldsymbol{s}_{m}}{\partial \Im\{\beta_{m'}\}}] = \delta_{mm'}
 \sum_{n=1}^N  e^{-jk_0 r_m} \boldsymbol{x}\otimes  [1,j],\\
\end{align}
where $A_{mni}$ and $c_{mi}$ are constants when $\boldsymbol{x}$, $\boldsymbol{q}$, and $\boldsymbol{p}_m$ are fixed. Thus, (\ref{FIM}) and (\ref{FIM2}) are only related to $\boldsymbol{w}$.

Denote $\mathbf{J}_f=\boldsymbol{J}_{\boldsymbol{q}\boldsymbol{q}}-\boldsymbol{J}_{\boldsymbol{q}\tilde{\boldsymbol{\gamma}}} \boldsymbol{J}_{\tilde{\boldsymbol{\gamma}}\tilde{\boldsymbol{\gamma}}}^{-1} \boldsymbol{J}_{\boldsymbol{q}\tilde{\boldsymbol{\gamma}}}^T$ .Then the mean square error (MSE) of any unbiased estimator $\hat{\boldsymbol{q}}$ of $\boldsymbol{q}$ is lower bounded by CRB 
\begin{align}
\text{CRB}(\boldsymbol{q}) = \operatorname{tr}\left\{\left[ \mathbf{J}_f \right]^{-1}\right\} \triangleq f(\boldsymbol{w}).
\end{align}
To enhance the positioning accuracy under the constraint on the RIS reflection coefficients, one needs to solve
\begin{align}
\mathcal{P}_1:  \min _{\boldsymbol{w}} \kern 3pt &f(\boldsymbol{w}) \\ \nonumber
\text {s.t. } &|w_n|=1, \kern 3pt n=1,\cdots,N.
\end{align}
Note that $f(\boldsymbol{w})$ is intrinsically influenced by EMI because the value of $P_m$ in (9) depends on EMI. 

Since both the objective function and the constraint in $\mathcal{P}_1$ are nonconvex,  $\mathcal{P}_1$ is non-convex and is challenging to solve. In the
following, an efficient algorithm is developed to derive a high quality optimal solution of $\boldsymbol{w}$.

\section{Manifold Optimization}
\subsection{Wirtinger Gradient}
Since $f(\boldsymbol{w})$ is a non-trivial (not constant) real-valued function with complex arguments $\boldsymbol{w}$, $f(\boldsymbol{w})$ is non-analytic and therefore is not complex differentiable. Thus, the steepest descent direction of $f(\boldsymbol{w})$ is $-\nabla_{\boldsymbol{w}^*}f(\boldsymbol{w})$ given by Wirtinger gradient \cite{Wirtinger}, where $\boldsymbol{w}^*$ denotes the conjugate of $\boldsymbol{w}$.
The $g$th element of $\nabla_{\boldsymbol{w}^*}f(\boldsymbol{w})$ is
\begin{align}
 \frac{\partial f(\boldsymbol{w})}
 {\partial w_g^*} &= - \operatorname{tr}[ \boldsymbol{J}_f^{-1}  \frac{\partial \boldsymbol{J}_f }
 {\partial w_g^*}
 \boldsymbol{J}_f^{-1} ] \nonumber\\
 & =- \operatorname{tr}[ \boldsymbol{J}_f^{-1}  \sum_{m=1}^M\frac{\partial  \boldsymbol{J}_{\boldsymbol{q}\boldsymbol{q},m} }
 {\partial w_g^*}
 \boldsymbol{J}_f^{-1} ].
\end{align}
According to \cite{Wirtinger}, the elements of $\frac{\partial\boldsymbol{J}_{\boldsymbol{q}\boldsymbol{q},m}}{\partial w_g^*}$ are computed by substituting (12) into (\ref{FIM}) as
\begin{align}
\left[\frac{\partial\boldsymbol{J}_{\boldsymbol{q}\boldsymbol{q},m}}{\partial w_g^*}\right]_{i, k}&=\frac{1}{P_m}
A_{mgk}^*\left(\sum_{n=1}^N w_n A_{mni}+c_{mi}\right)\nonumber\\
&+\frac{1}{P_m}A_{mgi}^*\left(\sum_{n=1}^N w_n A_{mnk}+c_{mk}\right)\nonumber\\
&-\frac{\left[\boldsymbol{J}_{\boldsymbol{q}\boldsymbol{q},m}\right]_{i, k}}{P_m}\left[\boldsymbol{w}^T \boldsymbol{H}_{m,2}^T \boldsymbol{R} \boldsymbol{H}_{m,2}^*\right]_{g}.
\label{18}
\end{align}
Note that the last term in (\ref{18}) represents the influence of EMI.

\subsection{Riemannian Gradient Descent}
The unit modulus constraint in $\mathcal{P}_1$ can be
geometrically interpreted as restricting $\boldsymbol{w}$  on the complex circle manifold that is  defined as
\begin{align}
\mathcal{M}=\left\{\boldsymbol{w} \in \mathbb{C}^N:\left|w_1\right|=\left|w_2\right|=\cdots=\left|w_N\right|=1\right\}.
\end{align}
The tangent space at $\boldsymbol{w}_i$ on the complex circle manifold is denoted as $T_{\boldsymbol{w}_i} \mathcal{M}$, which is the space of tangent vectors passing through $\boldsymbol{w}_i$ and is given by
\begin{align}
T_{\boldsymbol{w}_i} \mathcal{M}=\left\{\boldsymbol{v} \in \mathbb{C}^N: \Re\left(\boldsymbol{v} \odot \boldsymbol{w}_i^*\right)=\boldsymbol{0}_N\right\}.
\label{T}
\end{align}
Among all tangent vectors, the one that yields the fastest increase of the objective function
is defined as the Riemannian gradient, $\operatorname{grad}f(\boldsymbol{w}_i)$.
Numerically, $\operatorname{grad}f(\boldsymbol{w}_i)$ is computed  by first calculating the steepest ascent direction in the Euclidean space, i.e., $\nabla_{\boldsymbol{w}^*} f\left(\boldsymbol{w}_i\right)$, and then projecting it onto the tangent space via a projection operator. 
The projection operator from the Euclidean space onto the tangent space $T_{\boldsymbol{w}_i} \mathcal{M}$ is given by \cite{manopt}
\begin{align}
\mathcal{P}_{T_{\boldsymbol{w}_i} \mathcal{M}}(\boldsymbol{v})=\boldsymbol{v}-\Re\left\{\boldsymbol{v} \odot \boldsymbol{w}_i^*\right\} \odot \boldsymbol{w}_i .
\end{align}
Hence, the Riemannian gradient of $f\left(\boldsymbol{w}_i\right)$ is expressed as
\begin{align}
\operatorname{grad} f\left(\boldsymbol{w}_i\right) &=\mathcal{P}_{T_{\boldsymbol{w}_i} \mathcal{M}}\left(\nabla_{\boldsymbol{w}^*} f\left(\boldsymbol{w}_i\right)\right) \nonumber\\
&=\nabla_{\boldsymbol{w}^*} f\left(\boldsymbol{w}_i\right)-\Re\left\{\nabla_{\boldsymbol{w}^*} f\left(\boldsymbol{w}_i\right) \odot \boldsymbol{w}_i^*\right\} \odot \boldsymbol{w}_i.
\label{grad}
\end{align}  

We then employ the conjugate gradient (CG) method to find the search direction \cite{manopt}.
The update rule of the search direction in the Euclidean space
is given by 
\begin{equation}
\boldsymbol{\mu}_{i+1}=-\nabla_{\boldsymbol{w}_{i+1}^*} f+\alpha_i  \boldsymbol{\mu}_i,
\label{ser}
\end{equation}
where $\boldsymbol{\mu}_i$ denotes the search direction at $\boldsymbol{w}_{i}$ and $\alpha_i$ is chosen as the Polak-Ribiere parameter to achieve fast convergence
\cite{manopt}. However, since $\boldsymbol{\mu}_i$ and $\boldsymbol{\mu}_{i+1}$ in (\ref{ser}) lie in $T_{\boldsymbol{w}_i} \mathcal{M}$
and $T_{\boldsymbol{w}_{i+1}} \mathcal{M}$, respectively, they cannot be integrated directly
over different tangent spaces. Thus, we
need to project $\boldsymbol{\mu}_i$ from
tangent space $T_{\boldsymbol{w}_i} \mathcal{M}$ to tangent space $T_{\boldsymbol{w}_{i+1}} \mathcal{M}$ \cite{alternate}. 
Similar to (\ref{ser}), the search direction based on the Riemannian gradient can be updated as:
\begin{equation}
\boldsymbol{\mu}_{i+1}=-\operatorname{grad} f\left(\boldsymbol{w}_i\right)+\alpha_i \mathcal{P}_{T_{\boldsymbol{w}_{i+1}} \mathcal{M}}(\boldsymbol{\mu}_i),
\label{ser2}
\end{equation}
where the the Polak-Ribiere parameter $\alpha_i$ is computed as \cite{manopt}
\begin{equation}
\alpha_i=\frac{\operatorname{grad} f\left(\boldsymbol{w}_i\right)^H(\operatorname{grad} f\left(\boldsymbol{w}_i\right)-\operatorname{grad} f\left(\boldsymbol{w}_{i-1}\right))}{\operatorname{grad} f\left(\boldsymbol{w}_{i-1}\right)^H\operatorname{grad} f\left(\boldsymbol{w}_{i-1}\right)}.
\label{PR}
\end{equation}
However, given the search direction $\boldsymbol{\mu}_i$, the solution cannot be simply updated via $\boldsymbol{w}_{i+1}=\boldsymbol{w}_i+\eta_i\boldsymbol{\mu}_i$, where $\eta_i$ denotes the searching step size and $\eta_i \boldsymbol{\mu}_i$ is defined as $\textit{search vectors}$. The reason is that $\boldsymbol{w}_i+\eta_i \boldsymbol{\mu}_i$ would lie in the tangent space $T_{\boldsymbol{w}_i}\mathcal{M}$ but not on the surface of the manifold. 
Hence, a retraction function is needed from the tangent space to the surface of the manifold such that the RIS hardware limitation is not violated. For the complex circle manifold, the retraction function that maps $\boldsymbol{v}\in T_{\boldsymbol{w}_i} \mathcal{M}$ onto $\mathcal{M}$ can be defined as \cite{manopt}
\begin{align}
\mathcal{R}_{\boldsymbol{w}_i}(\boldsymbol{v})=(\boldsymbol{w}_i+\boldsymbol{v}) \oslash |\boldsymbol{w}_i+\boldsymbol{v}|,
\label{retr1}
\end{align}
where $\oslash$ denotes element-wise division. We then adopt retraction to find the next iterate $\boldsymbol{w}_{i+1}$ on the
manifold as
\begin{align}
\boldsymbol{w}_{i+1}=\mathcal{R}_{\boldsymbol{w}_i}(\eta_i \boldsymbol{\mu}_i),
\label{retr}
\end{align}
where the Armijo backtracking line search algorithm \cite{manopt} can be used to choose the 
step size $\eta_i$, which ensures that the objective function is decreasing at each iteration, i.e., $f(\boldsymbol{w}_{i+1}) < f(\boldsymbol{w}_{i})$.

A locally optimal solution can be found by repeating the above steps until $\|\operatorname{grad}f(\boldsymbol{w}_{k}))\|_2<\epsilon$, where $\epsilon$ is the convergence tolerance, and $k$ is the maximum number of iteration until convergence. This procedure is defined as Riemannian gradient descent (RGD). 

\subsection{Riemannian Nonlinear Acceleration}
Although the monotonic decrease in the objective function is guaranteed after each iteration, the convergence is reached with maximum number of iteration $k=\mathcal{O}(\epsilon^{-2})$ \cite{conver}.
When small convergence tolerance is demanded, $k$ may be unacceptably large, and the RGD algorithm may be slow in practice. Hence, we apply the regularized nonlinear acceleration gradient algorithm \cite{NA} to accelerate the RGD algorithm in parallel.

Let us consider a generalization of nonlinear acceleration for Riemannian optimization via a weighted Riemannian average on the manifold. Define $M_d$ as the memory depth that controls the interval between two acceleration operations.
After obtaining a sequence of iterates from the RGD process, denoted as  $\left\{\mathbf{w}_i\right\}_{i=0}^{M_d}$, we define the residuals as 
the projection of all search vectors  onto the tangent space at $\mathbf{w}_{M_d-1}$, which is expressed as
\begin{align}
\mathbf{r}_i= \mathcal{P}_{T_{\mathbf{w}_{M_d-1}} \mathcal{M}} \left( -\eta_i \operatorname{grad}f(\mathbf{w}_i)\right).
\label{r}
\end{align}
The weights for the acceleration is defined as $\mathbf{c}$, and can be obtained by minimizing the sum of a weighted combination of the
residuals and a regularization term in the weights \cite{NA}. The solution of $\mathbf{c}$ is given by the following optimization problem 
\begin{align}
\mathbf{c}=\mathop{\arg\min}_{\mathbf{c} \in \mathbb{R}^{M_d}: \mathbf{c}^{\top} \mathbf{1}=1}   \left\{\left\|\sum_{i=0}^k c_i \mathbf{r}_i\right\|_2^2+\lambda\|\mathbf{c}\|_2^2\right\},
\label{c}
\end{align}
where  $\mathbf{1}$ denotes a vector with all elements equal to $1$, and $\lambda$ is the regularization parameter.
We then show that the optimal weight $\mathbf{c}$ has a closed-form solution.
Define the residual matrix as $\mathbf{R}=\left[ \Re\{\mathbf{r}_i^H \mathbf{r}_j \} \right]_{i, j} \in \mathbb{C}^{M_d \times M_d}$, which collects all pairwise inner products of $\mathbf{r}_i$. Note that (\ref{c}) is a linearly constrained quadratic program that can be solved by introducing  
 $\mu \in \mathbb{R}$ as the dual variable. Then (\ref{c}) indicates that $\mathbf{c}$ and $\mu$ satisfy the Karush-Kuhn-Tucker (KKT) system:
\begin{align}
\left[\begin{array}{cc}
2(\mathbf{R}+\lambda \mathbf{I}) & \mathbf{1} \\
\mathbf{1}^{\top} & 0
\end{array}\right]\left[\begin{array}{l}
\mathbf{c} \\
\mu
\end{array}\right]=\left[\begin{array}{l}
\mathbf{0} \\
1
\end{array}\right].
\label{KKT}
\end{align}
Solving (\ref{KKT}) yields the closed form of $\mathbf{c}$ as
\begin{align}
\mathbf{c}=\frac{(\mathbf{R}+\lambda \mathbf{I})^{-1} \mathbf{1}}{\mathbf{1}^{\top}(\mathbf{R}+\lambda \mathbf{I})^{-1} \mathbf{1}},
\label{C}
\end{align}
where $\mathbf{I}$ is the identical matrix.
The sequence of converging iterates from the RGD process recursively constructs an updated iterates $\tilde{\mathbf{w}}_i (i=0,\cdots,M_d-1)$
as
\begin{align}
\tilde{\mathbf{w}}_i=\mathcal{R}_{\tilde{\mathbf{w}}_{i-1}}\left(\frac{c_i}{\sum_{j=0}^i c_j} \mathcal{R}_{\tilde{\mathbf{w}}_{i-1}}^{-1}\left(\mathbf{w}_i\right)\right),\tilde{\mathbf{w}}_{-1}=\mathbf{w}_{0}.
\label{new}
\end{align}
The inverse of the retraction $\mathbf{v}=\mathcal{R}_{\mathbf{w}_{i}}^{-1}(\mathbf{w}_{i+1})$ maps $\mathbf{w}_{i+1} \in \mathcal{M}$ to $\mathbf{v} \in T_{\mathbf{w}_{i}}\mathcal{M}$, and can be derived as follows. Since the angles of all elements in $(\mathbf{w}_i+\mathbf{v})$ are unchanged in (\ref{retr}), there is 
\begin{align}
\operatorname{arg}(\mathbf{\mathbf{w}}_i+\mathbf{v})=\operatorname{arg}(\mathbf{\mathbf{w}}_{i+1}),
\label{C1}
\end{align}
where $\operatorname{arg}(\cdot)$ denotes the arguments of all elements in a vector.
Because of $\mathbf{v} \in T_{\mathbf{w}_{i}}\mathcal{M}$, (\ref{T}) can be rewritten as 
\begin{align}
\operatorname{arg}(\mathbf{v})-\operatorname{arg}(\mathbf{w}_i)=\pm \frac{\pi}{2}.
\label{C2}
\end{align}
Based on the geometrical constraints in (\ref{C1}) and (\ref{C2}), the closed form of the inverse of the retraction is given by
\begin{align}
\mathcal{R}_{\mathbf{w}_{i}}^{-1}(\mathbf{w}_{i+1})=j \mathbf{w}_{i} \odot \tan \left(\operatorname{arg}(\mathbf{w}_{i+1})-\operatorname{arg}(\mathbf{w}_{i})\right).
\label{inv}
\end{align}
Substituting (\ref{inv}) into (\ref{new}), the iterative sequence is updated and RGD can restart with $\mathbf{w}_{0}=\tilde{\mathbf{w}}_{M_d-1}$. 
The nonlinear acceleration is performed in parallel to the RGD process in the complete procedure, which is summarized in Algorithm 1.
 
\begin{algorithm}[t]  
  \caption{Riemannian gradient descent (RGD) with nonlinear acceleration}  
  \label{alg:Framwork}  
  \begin{algorithmic} [1]
    \Require  
    Initialization $\mathbf{w}_0$, regularization parameter $\lambda$, convergence tolerance $\epsilon$, and memory depth $M_d$. 
    \While{$\|\operatorname{grad}f(\mathbf{w}_{0}))\|_2 \ge \epsilon$}
    \For{$i=1$ to $M_d$}
       \State Compute $\operatorname{grad} f\left(\mathbf{w}_i\right)$ via (\ref{grad}).
       \State Compute $\boldsymbol{\mu}_{i-1}$ via (\ref{ser2})
       with $\boldsymbol{\mu}_{-1}=\mathbf{0}$.
       \State Update $\mathbf{w}_i=\mathcal{R}_{\mathbf{w}_{i-1}}(\eta_{i-1} \boldsymbol{\mu}_{i-1})$ with $\eta_{i-1}$ given by Armijo backtracking line search [16, 4.2.2].
  	\EndFor 
  	\State Compute $\mathbf{r}_i (i=0,\cdots,M_d-1)$ via (\ref{r}).
  	\State Compute $\mathbf{c}$ via (\ref{C}).
  	\State Compute $\tilde{\mathbf{w}}_i (i=0,\cdots,M_d-1)$ via (\ref{new}), with $\tilde{\mathbf{w}}_{-1}=\quad \quad \mathbf{w}_{0}$.
  	\State Restart with $\mathbf{w}_{0}=\tilde{\mathbf{w}}_{M_d-1}$.
  	\EndWhile
  \end{algorithmic}  
\end{algorithm}

\section{Simulation Results and Analysis}
In the simulations, three anchors are located at $\boldsymbol{p}_1=(-20,50,30)$m, $\boldsymbol{p}_2=(-15,35,40)$m, and $\boldsymbol{p}_3=(-15,60,35)$m, respectively, and we consider three different values of frequencies of the transmitted signal as $30$GHz. The length of each element on the RIS is set as $l_1=l_2=0.01$m, and the spacing distance between two elements is set as $0.01$m. The EMI is assumed uniformly distributed from all angles with $E_{EMI}=-70$dBW/$m^2$. Moreover, we set $\sigma_m^2=-124$dBW.
In the RGD process, we choose $\lambda=10^{-7}, \epsilon=10^{-3}$, and $M_d=5$ and randomly set the initial value of $\boldsymbol{w}$ as $\boldsymbol{w}_0$.
The root of Cram\'{e}r-Rao Bound (RCRB) is selected as the criterion to measure the localization accuracy.






\subsection{The Riemannian Gradient Descent Process}
\begin{figure}[t]
  \centering
  \centerline{\includegraphics[width=8cm,height=5cm]{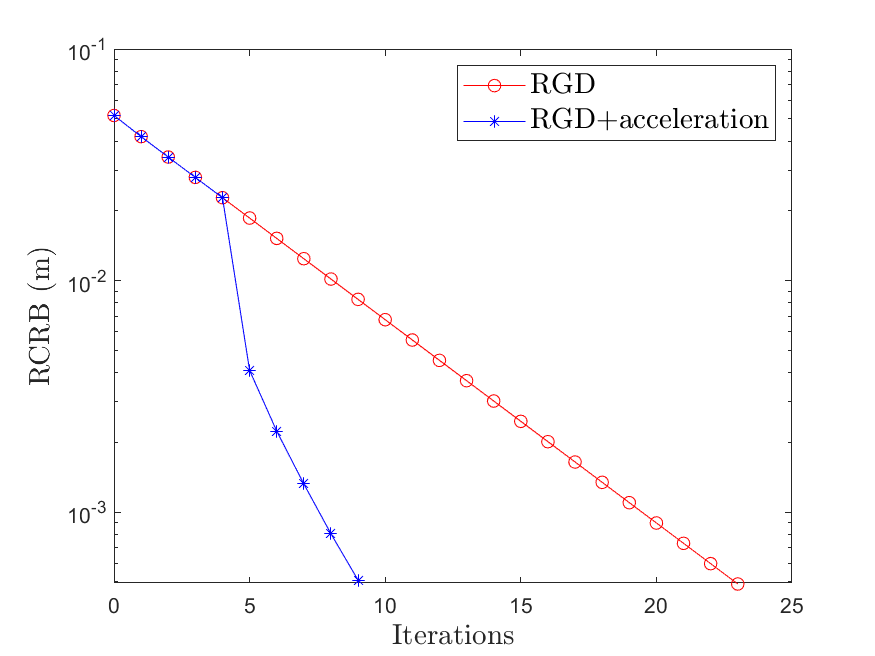}}
  \caption{Riemannian gradient descent algorithm with and without Riemannian nonlinear acceleration.}
  \label{P1}
\end{figure}
The position of the agent is set at $\boldsymbol{q}=(0,0,20)$m, and the length of the RIS is set as $a=b=0.8$m. We perform the Riemannian gradient descent algorithm with and without Riemannian nonlinear acceleration in Fig.~\ref{P1}.
It is seen that the acceleration technique can reduce 
the iteration number more than half when the same RCRB is reached.

\subsection{The RCRB versus The Distance from The Agent to The RIS}
\begin{figure}[t]
  \centering
  \centerline{\includegraphics[width=8cm,height=5cm]{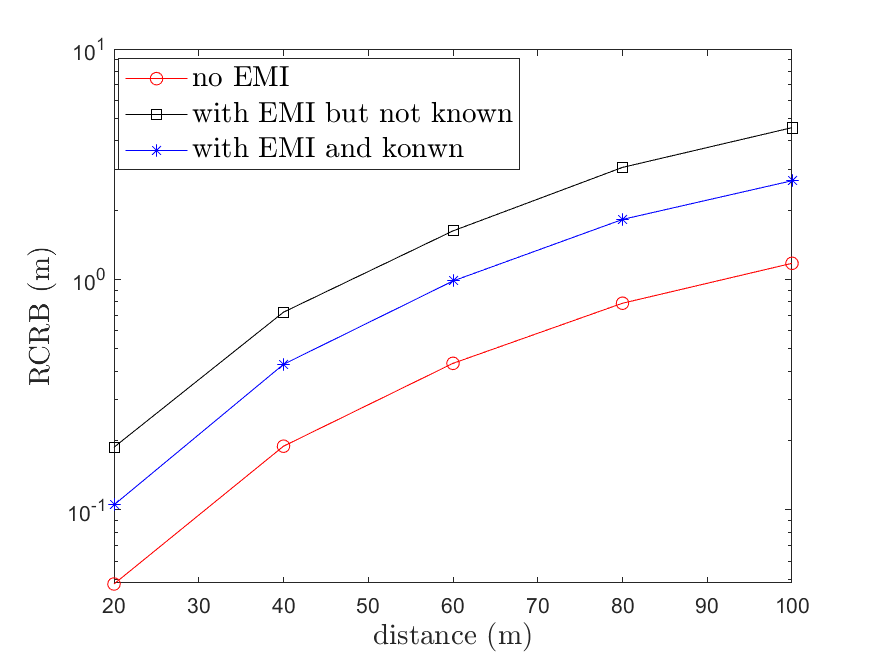}}
  \caption{The RCRB versus the distance from the agent to the RIS. }
  \label{P2}
\end{figure}
The position of the agent is set at $\boldsymbol{q}=(0,0,d)$, where $d$ denotes the distance from the agent to the RIS, and the length of the RIS is set as $a=b=0.6$m.
We plot the change of RCRB when the agent becomes farther from the RIS in Fig.~\ref{P2}.
It is seen that RCRB becomes larger with the increase of the distance, because when the source becomes farther from the RIS, the power illuminated on the RIS becomes less and  the amplitude of the signal received by the anchors becomes smaller.
In the presence of unknown EMI, the optimization is performed when assuming  $P_m=\sigma_m^2$ in (7). Since $P_m$ is actually enlarged by EMI, the CRB becomes drastically larger. When the statistical information of EMI is known, the proposed method can alleviate the CRB degradation caused by EMI.

\subsection{The RCRB versus The Length of The RIS}
\begin{figure}[t]
  \centering
  \centerline{\includegraphics[width=8cm,height=5cm]{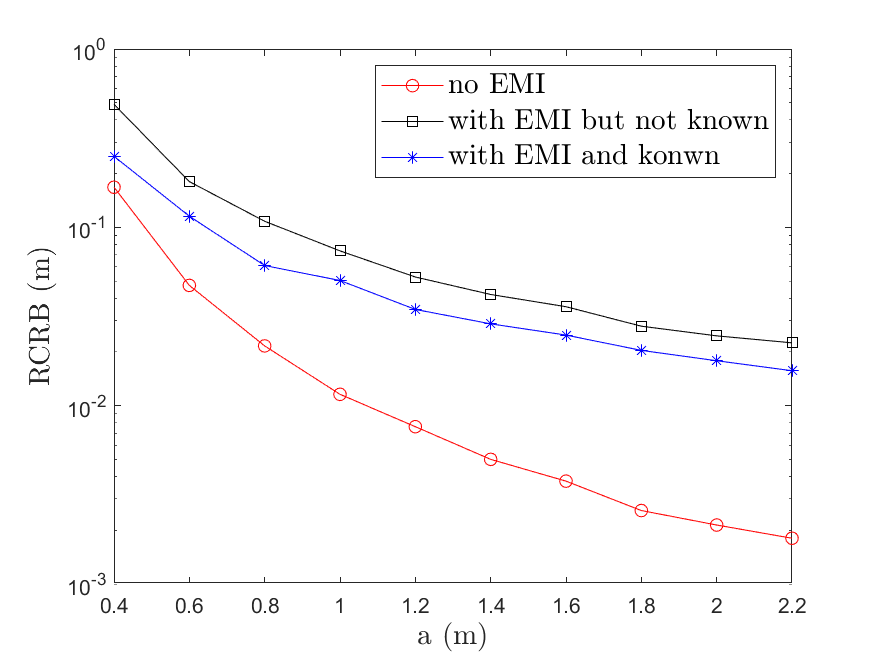}}
  \caption{The RCRB versus the length of the RIS.}
  \label{P3}
\end{figure} 
We set the position of the agent at $\boldsymbol{q}=(0,0,20)$m, and set $a=b$ for the  RIS in squared shape.
We plot the change of RCRB when the length of the RIS becomes larger in Fig.~\ref{P3}.
It is seen that RCRB becomes smaller with the increased length of the RIS. When the length of the RIS becomes larger, the positioning accuracy loss caused by EMI becomes larger, because the general EMI power captured by RIS becomes higher, while the proposed method can alleviate the positioning accuracy loss regardless of the length of the RIS.



\section{Conclusion}
In this paper, we apply the manifold optimization method to derive the locally optimal CRB of the localization error under EMI and with practical RIS hardware limitations, where the Wirtinger gradient is calculated to find the iterative search direction. 
To solve the problem of slow convergence, the Riemannian nonlinear acceleration technique that speeds up the convergence rate is employed.
Simulation results show that the proposed method can significantly decrease the CRB of the localization error when EMI degrades the localization accuracy.


 \small 
 \bibliographystyle{ieeetr}
 \bibliography{IEEEabrv,mylib}


\ifCLASSOPTIONcaptionsoff
  \newpage
\fi



%

\end{document}